\documentclass[aps,twocolumn,showpacs,preprintnumbers,prl,superscriptaddress,groupedaddress,10pt]{revtex4-1}
\usepackage[utf8]{inputenc}
\usepackage{graphicx,amssymb,amsmath,amsthm,amsfonts,epsfig,times,natbib}

\usepackage[usenames,dvipsnames]{color}
\usepackage{epstopdf}
\usepackage{amsmath,amssymb}
\usepackage{tensor}
\usepackage{mathtools}
\usepackage{amsbsy}
\usepackage{bm,url}
\usepackage[linktocpage]{hyperref}

\begin{document}


\textcolor{blue}{}

\textcolor{red}{}




\title{On the formation of "supermassive" neutron stars and dynamical transition to spontaneous scalarization}

\author{Juan Carlos Degollado$^{1}$}
\email{jcdegollado@icf.unam.mx}

\author{Marcelo Salgado$^{2}$}
\email{marcelo@nucleares.unam.mx}

\author{Miguel Alcubierre$^{2}$}
\email{malcubi@nucleares.unam.mx}

\affiliation{$^1$ Instituto de Ciencias F\'isicas, Universidad Nacional Aut\'onoma 
de M\'exico, Apartado. Postal 48-3, 62210, Cuernavaca, Morelos, M\'exico. \\
$^2$Instituto de Ciencias Nucleares, Universidad Nacional Aut\'{o}noma de M\'{e}xico,
 A.P. 70-543, M\'{e}xico D.F. 04510, M\'{e}xico.}


\date{\today}

\begin{abstract}
It is well known that neutron stars can undergo a phase transition under a certain class of 
Scalar Tensor Theories of gravity (STT's) where a new order parameter, the {\it scalar charge}, 
appears within the star. This is the well known phenomenon of spontaneous scalarization (SC) discovered 
by Damour and Esposito-Far\`ese in 1993. Under such mechanism neutron stars can afford in principle a maximum 
mass larger than in general relativity (GR) for a given equation of state without taking into account additional observational constraints (e.g. binary systems). This opens the possibility 
that neutron stars might be formed with masses as large as 
$\sim 2 M_\odot$ without the need of stiff, or more exotic, equations of state for the nuclear matter. Thus, STT's through SC may account for compact objects with large masses observed recently in the sky in the form of pulsars (PSR J0348+0432 with 
$M= 2.01 M_\pm 0.04\odot$ observed in 2013, PSR J1614-2230 with $M= 1.97\pm 0.04  M_\odot$ observed in 2010 or J0740+6620 $M= 2.14^{+0.10}_{-0.09} M_\odot$ observed in 2019). However, we argue that even if that was possible such maximum mass models within STT cannot be 
formed solely from the dynamic transition of an initial ``isolated'' unscalarized neutron star whose mass cannot exceed the maximum mass in GR. This is because SC, being an energetic-preferred configuration, produces a final static star with a mass lower that the initial one with a fixed baryon mass. The mass difference between the 
initial and final configurations is radiated away in the form of a scalar-field. Thus, maximum mass models of scalarized neutron stars, if present in nature, must have formed by a different process, perhaps of cosmological origin or by the subsequent accretion of additional scalar charge and mass.
\end{abstract}



\maketitle

{\bf Introduction.}
Relativistic theories of gravity different from general relativity (GR) \cite{Will2018,Will:2005va,Berti:2015itd, Baker:2019gxo, Burrage2018}
have received a major interest in recent years for various reasons. 
For instance, several phenomena in our universe can be explained by the presence of some unknown form of matter and energy, dubbed 
{\it dark}~\cite{Joyce:2014kja}. The anisotropies of temperature variations in the cosmic microwave background, the flat rotation curves of galaxies, different clustering properties of large scale structures in the cosmos, and the accelerated expansion in the universe, are perhaps the most conspicuous phenomena that require dark components. 
However, the failure of a direct detection of some of these dark substances by several dark-matter laboratories, as well 
as the unsuccessful efforts to accommodate the cosmological constant (the simplest form of dark energy) in a consistent manner within the framework of a robust quantum field theory (see~\cite{Martin2012} for a thorough review), have led to an alternate direction consisting in modifying the gravitational sector, i.e., modifying GR~\cite{Koutsoumbas2018,Faraoni04,Tsujikawa:2008,DeFelice:2010b,Khoury2013, Burrage:2017qrf}.
Nonetheless, the amazing success of GR at different scales makes also difficult for alternative theories of gravity (ATG) to reproduce in a fully consistent fashion the standard tests of GR while providing the required features without dark sectors.
Scalar tensor theories of gravity (STT) in its simplest form (see below)~\cite{Damour92,Riazuelo2002,Fujii2003,Faraoni04} are one of the few ATG that remains as a viable candidate to generalize 
Einstein's GR while making clear cut predictions that can be falsified by several kind of experiments. At cosmological level, 
they can produce an EOS for the dark energy (in the form of a scalar-field) which can vary in cosmic time, unlike the cosmological constant $\Lambda$. This variation will be tested in the forthcoming years by numerous projects~\cite{DESI,LSST,Amendola:2018,EUCLIDES} which will help to validate or rule out a simple $\Lambda$CDM model.
Since STT are a generalization of the famous Brans-Dicke theory, these theories predict small variations of the effective gravitational coupling $G_{\rm eff}$, which can be further tested observationally.
Furthermore, STT predicts the existence of an additional polarization mode of gravitational waves, the {\it breathing mode}, due to the presence of a scalar-field that is non-minimally coupled (NMC) to gravity which can 
be detected or constrained in the future~\cite{Abbott:2019}. This type of polarization mode can be revealed itself in the strong gravity regime even in spherical symmetry, unlike the usual tensor polarizations ($h_{+}, h_{\times}$) that require the presence of sources with relatively large quadrupolar time variations. The natural place to look for that kind of gravitational radiation is in neutron stars, notably in pulsars.
It is well known since the early 1990's, from the pioneer analysis by Damour and Esposito-Far\'ese (DEF)~\cite{Damour93,Damour96} that neutron stars can undergo a phase transition in the form of {\it spontaneous scalarization} (SC) where a new order parameter, dubbed {\it scalar charge}, can appear within the star if the latter is sufficiently compact. Therefore, at the threshold of this transition corresponding to a critical central energy-density (or alternatively a critical total baryon mass) the star transits to a configuration of lower energy (i.e. lower gravitational mass) while keeping the total number of baryons fixed. The difference of gravitational mass between the initial (unstable) configuration and the final static scalarized configuration is radiated away in the form of scalar gravitational waves of the sort described above.
Following the DEF discovery, subsequent analyses showed that SC is robust in that it is independent of the equation of state (EOS) for the nuclear matter adopted to model a neutron star~\cite{Salgado98,Novak98b,Damour98,Anderson2019}. Finally, a notable feature of SC is that maximum mass models of neutron stars constructed within some specific classes of SST can be much larger than the corresponding models in GR while keeping fixed the EOS~\cite{Damour93,Salgado98}. The recent  observations of massive neutrons stars in the form of pulsars (PSR J0348+0432 with 
$M= 2.01 \pm 0.04 M_\odot$~\cite{Antoniadis2013}, PSR J1614-2230 with $M= 1.97\pm 0.04  M_\odot$~\cite{Demorest2010} or J0740+6620 $M= 2.14^{+0.10}_{-0.09} M_\odot$~\cite{Cromartie2019}; 
if confirmed PSR J2215+5135 might host the largest neutron-star mass observed to date $M= 2.27^{+0.17}_{-0.15} M_\odot$~\cite{Linares2018}) put stringent constraints on the current EOS 
of nuclear matter used to model neutron stars in GR~\cite{Chamel2013,Zhou2019,Salgado94}. This opens the door 
for the large observed masses of neutron stars 
to be explained by the phenomenon of SC in STT without the requirement of very stiff or more exotic EOS as it would be the case if one works under the framework of pure GR . 

Nevertheless, and this is the most important contribution of this letter, we argue that if such massive neutron stars are explained by the SC transition, the latter must be accompanied by an additional process, maybe of cosmological origin or by accretion, because SC, at least in its standard conception, requires that the initial configuration (one which coincides with a static configuration in GR) has a mass that cannot be larger that the maximum mass allowed by the unscalarized neutron star which maybe lower than $2M_\odot$ if the EOS considered is not very stiff \cite{Chamel2013,Zhou2019}. In order to support this conclusion, we perform a thorough numerical analysis by evolving numerically the full non-linear system of equations in STT in the {\it Jordan frame} (as opposed to the Einstein frame which is the most frequently used~\cite{Novak98b,Ortiz2016}), including the relativistic hydrodynamical sector that represents the star (modeled by a perfect fluid), under the assumption of spherical symmetry. As initial data we select a large sample of possible initial configurations of unscalarized neutron stars in hydrostatic equilibrium. By definition, the initial data corresponds to an initial data in GR where the scalar-field is null. Near the threshold of instability towards SC the initial configuration is perturbed by a Gaussian scalar-field profile, and then a numerical evolution is performed until reaching the final state of the system which corresponds to a static scalarized neutron star with gravitational mass lower than the initial one but with the same initial baryon mass.
This analysis has several consequences, both theoretical and observational. From the theoretical point of view it seems impossible that a dynamical process like the one described here may lead by its own to a scalarized static star with a gravitational mass larger than the initial one since by definition, {\it spontaneous scalarization} corresponds to a static configuration with energy (i.e. gravitational mass) lower than the energy in GR while keeping the total baryon mass fixed (see Fig.1 in \cite{Damour:1993hw}, and Fig. 6 in \cite{Salgado98}). That is, a scalarized neutron star is the energetically preferred configuration above a critical energy-density. As we stressed above, scalarized configurations with gravitational masses larger than the maximum mass in GR were predicted in the past by constructing spherically symmetric configurations by solving directly the field equations of STT under the assumption of strict staticity~\cite{Damour:1993hw,Salgado98}. Thus, the scalarized neutron star models with masses larger than the maximum mass in GR, with fixed EOS, cannot be the result from the purely dynamical transition of initial lower mass configurations. {\it A fortiori} such massive scalarized stars must be formed dynamically by a more complicated process that is worth exploring in the future. From the observational point of view, it is still unknown if (large or low mass) scalarized neutron stars really exist in nature. On one hand, detailed observations in binary pulsars put constraints on the possible values of the NMC between the scalar-field and the Ricci curvature. For certain class of STT, the weaker the NMC the lower the scalar charge, in which case,  dynamical SC might not be a viable mechanism to explain the recently detected massive neutron stars without appealing to stiff or to more exotic EOS. On the other hand, the current observations of gravitational waves by the LIGO-VIRGO collaboration, notably, by the event GW170817 involving the collision of two neutron stars~\cite{Abbott:2017a}, together with the 
simultaneous detection of gamma-ray bursts~\cite{Abbott:2017b} have allowed to rule out several ATG~\cite{Ezquiaga:2017,Baker:2017,Sakstein:2017}. Notwithstanding, such event did not have enough precision to constraint by its own the existence of an additional polarization mode, like the breathing scalar mode alluded above. One can avoid, in principle, the constraints by the binary pulsars in STT by including a mass term for the scalar field~\cite{Pretorius2016} as in this case, the field becomes of shorter range depending on the magnitude of the mass.
In the following sections we describe the formalism and the numerical results on the dynamic transition to SC using {\it ab initio} the Jordan frame where the physics is better understood.
We conclude the paper by discussing several directions of study like considering a mass term and speculate also about possible scenarios leading to the formation of massive scalarized neutron stars other that the dynamical transition explored here.

{\bf Scalar Tensor Theories and Spontaneous Scalarization.}
We consider the action for STT in the Jordan frame given by
\begin{eqnarray}
S[g_{ab},\phi] &=&
\int \left[\frac{f(\phi)}{2} R
- \frac{1}{2}(\nabla \phi)^2 - V(\phi)\right] \sqrt{-g} \: d^4x
\nonumber\\
&+& S_{\rm matt} \; ,
\label{eq:action_jordan}
\end{eqnarray}
where $S_{\rm matt}$ represents the action for the matter part that in the present case corresponds to a perfect fluid, $f(\phi)$ is
a NMC function 
$f(\phi) = \frac{1}{\kappa}(1 +  \kappa \xi \phi^2) \,$,
with $\kappa=8\pi G_0$, $G_0$ is Newton's gravitational constant, $\xi$ a positive dimensionless constant, and $V(\phi)$ is a scalar potential. Variation of the action~(\ref{eq:action_jordan}) with respect to the metric and 
with respect to the scalar field provide, respectively, the following two field equations:
$R_{ab}-\frac{1}{2}g_{ab}R = \kappa T_{ab} $ and $g^{ab}\nabla_a\nabla_b \phi + \frac{1}{2}f^\prime R = V^\prime\,,
$
%
%
where a prime indicates derivatives with respect to the 
scalar field. The effective energy-momentum tensor (EMT) is given by 
the contribution of three parts:
$T_{ab} := \left(T_{ab}^f
+ T_{ab}^{\phi} + T_{ab}^{{\rm fluid}}\right)/(\kappa f)$, where
\begin{align}  
T_{ab}^f &:= \nabla_a \left( f^\prime 
\nabla_b\phi\right) - g_{ab}\nabla_c \left(f^\prime 
\nabla^c \phi\right) \; ,
\label{eq:TabF} \\
T_{ab}^{\phi} &:=  (\nabla_a \phi)(\nabla_b \phi) - g_{ab}
\left[ \frac{1}{2}(\nabla \phi)^2 + V(\phi) \right ] \; , \quad
\label{eq:Tabphi}\\
T_{ab}^{\rm fluid}&:=
 (\mu  + p/c^2) u_{a}u_{b} + pg_{ab}\ ,
\label{eq:Tabfluid}
\end{align}
$\mu$ stands for the {\it total} mass density of matter in the 
rest frame of the fluid, and $p$ is the pressure as measured in the same frame. The diffeomorphism invariance of STT leads to the conservation of the EMT of matter alone: $\nabla_a T^{ab}_{\rm fluid}=0$, which in turn provides the hydrodynamic equations for the fluid. 
For simplicity we take $V(\phi)\equiv 0$ (i.e., a massless and non-self interacting scalar field) like in the original SC scenario~\cite{Damour93,Salgado98}. Notwithstanding, the massive case has been analyzed recently in static situations as a mechanism to avoid the constraints imposed by pulsars~\cite{Pretorius2016}. In the future we plan to study the dynamical transition to SC by taking into account a mass term as well. This paper 
focuses on solving the field equations of STT under the assumption of spherical symmetry. The details concerning the 
system of equations under this symmetry will be reported elsewhere. The interested reader is urged to consult \cite{Ruiz:2012jt} in order to have a glimpse of this system, but suffice is to say that in this analysis we use radial coordinates of area-type as opposed to isotropic. In~\cite{Ruiz:2012jt} we considered a complex-valued boson field (i.e. a boson star) as matter instead of a perfect fluid. 
As we mentioned before, the dynamical transition to SC have been analyzed by several authors in the past, but in 
the Einstein frame~\cite{Novak98b,Ortiz2016} where the field is coupled minimally to the 
curvature but non-minimally to the matter sector. 

{\bf Stellar models.} 
We consider a neutron-star model that initially is in hydrostatic 
equilibrium. The conservation of the EMT leads to an
equation which is formally identical to the Tolman-Oppenheimer-Volkoff (TOV) equation, except that the metric components take a different form due to the contribution of the scalar-field. For simplicity and in order to avoid interpolation of data of realistic EOS, we model the internal structure by a simple polytropic EOS 
parametrized by the baryon density $n_b$
:
\begin{align}  
 \frac{p}{c^2} &= \bar \kappa m_b n_0 \left(\frac{n_b}{n_0}\right)^{\gamma} \ ,\\
 \mu &= m_b n_b + \frac{p/c^2}{\gamma-1} \ ,
\end{align}
where $m_b =1.66 \times 10^{-24}$\;g and $n_{0}=0.1$ fm$^{-3}$.
The values of the parameters $\gamma$ and $\bar\kappa$ are adjusted to fit a stiff and a soft EOS. In particular, we use $\gamma=(2.34,\,2.46)$, and $\bar\kappa=(0.0195,\, 0.0093)$  like in Ref.~\cite{Damour:1993hw}.


Given this explicit EOS, we solved the TOV equation and constructed an initial static configuration where the star is not yet scalarized, and thus, the scalar field is absent. Thus, initially all the equations reduce to the field equations in GR. However, as the scalar-field evolves, automatically the field equations of STT are taken fully into account. In order to trigger easily the dynamical transition to SC the initial static configuration must be near the threshold of instability towards SC and then add a small Gaussian perturbation to the scalar field. If we departed from a static configuration not very near that threshold, 
while keeping the perturbation small, the latter is simply radiated away and the star is not scalarized. Under those initial conditions the star evolves 
towards the scalarized state, and the scalar field grows at the center of the star until the star stabilizes again into a different "hydrostatic" equilibrium configuration with a lower central energy density. Part of the total energy is radiated away in the form of scalar gravitational waves, and during this process the star develops a new global quantity, dubbed the 
{\it scalar charge} $\omega$, defined asymptotically in terms of a surface integral as follows:
\begin{equation}
\omega:= -\lim_{r \rightarrow \infty} \frac{1}{4\pi\,\sqrt{G_0}}
\int_{S} s_a \nabla^a\,\phi\,ds \; .
\label{eq:scalarQ}
\end{equation}
The factor $1/4\pi\,\sqrt{G_0}$ as well as the minus sign are a matter of convention, and $\omega$ has mass units. Since at the end of the transition the scalar field behaves asymptotically as $\phi\sim \sqrt{G_0} \omega/r$, we can simply extract $\omega$ from this expression in the asymptotic region.  

\begin{figure}[!ht]
\includegraphics[width=0.5\textwidth, height=0.2\textheight]{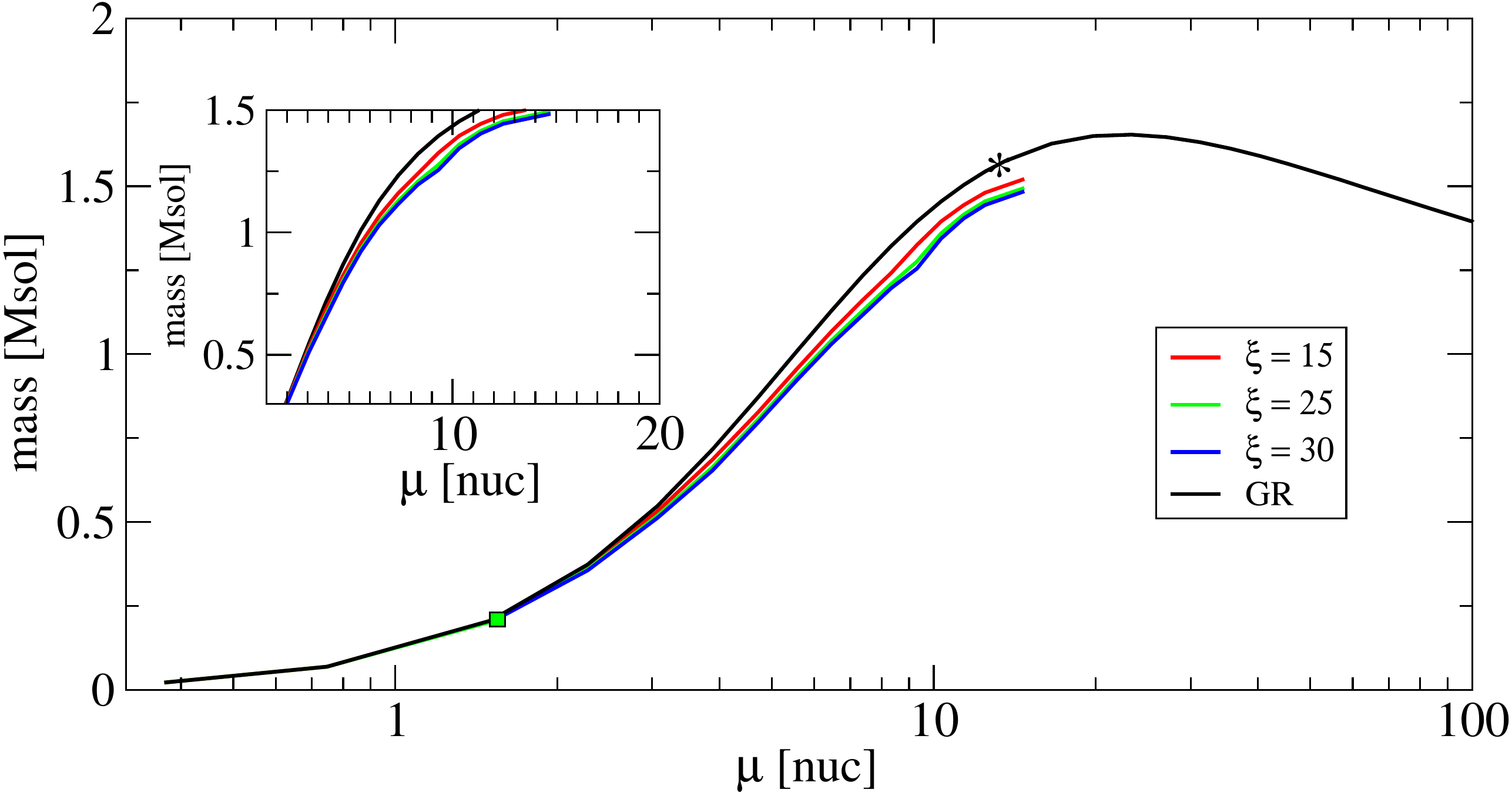}
\caption{Neutron star mass (in solar mass units) as a function of the central value of
the (mass) density for a few sequences of equilibrium models in GR and for a STT with $\xi=15,25,30$.  The maximum mass in GR, separates the
stable and unstable configurations towards black hole formation. The units used are in terms of the nuclear density $\mu_{\rm nuc}=1.66 \times 10^{17}$ kg/m$^3$. 
For small values of $\mu$ the curves overlap. The * indicates the configuration associated with Figs. \ref{fig:metric} and \ref{fig:matter}.
}
\label{fig:k1527_mu_vs_mass_xis_inset}
\end{figure}

We choose $\xi\geq 15$ in order to 
enhance the appearing of SC
\footnote{In Sec.~VI of\cite{Damour96} a STT similar to the one presented here is considered. In order to match notations their scalar field $\Phi$ and ours $\phi$ are related by $\Phi=\sqrt{\kappa}\phi$. In which case, the NMC constant $\xi$ coincide.}. However, according to~\cite{Damour96,Freire2012} the {\it curvature parameter} $\beta_0=-2\xi$ associated with STT 
is such that $-5\lesssim\beta_0$ 
in order to satisfy the constraints imposed by several pulsars. This constraint translates into $\xi\lesssim 2.5$.
Intriguingly, for sufficiently high $\xi$ the star properties become {\it universal} (cf. Figure~\ref{fig:k1527_mu_vs_mass_xis_inset}). For values 
$0<\xi\lesssim 15$ with the polytropic EOS described before, SC is basically unobserved. That is, the resulting star configurations are practically the same as in GR.

{\bf Maximal masses.}
The mass of a neutron 
star is computed using the definition of the Misner-Sharp
mass function, $M _{MS}= \frac{r_{out}}{2}\left(1 - \frac{1}{g_{rr}} \right)$ where $r_{out}$ is the radial coordinate at the outer boundary of the numerical domain. This mass is basically the ADM mass $M_{\rm ADM}$ minus the energy radiated away in the form of scalar gravitational waves. At the end of the evolution $M _{MS} < M_{\rm ADM}$, but initially, $M_{MS}^{t=0}=M_{\rm ADM}$. We checked that during the evolution the total baryon mass $M _{\rm bar}$ remains the same, corroborating that the preferred static configuration with fixed $M _{\rm bar}$ is the one with lower gravitational mass $M _{MS}^{t=t_f}< M_{MS}^{t=0}$, here $t_f$ 
stands for the end of the transition to SC.

\begin{figure}[!ht]
\includegraphics[width=0.49\textwidth, height=0.35\textheight]{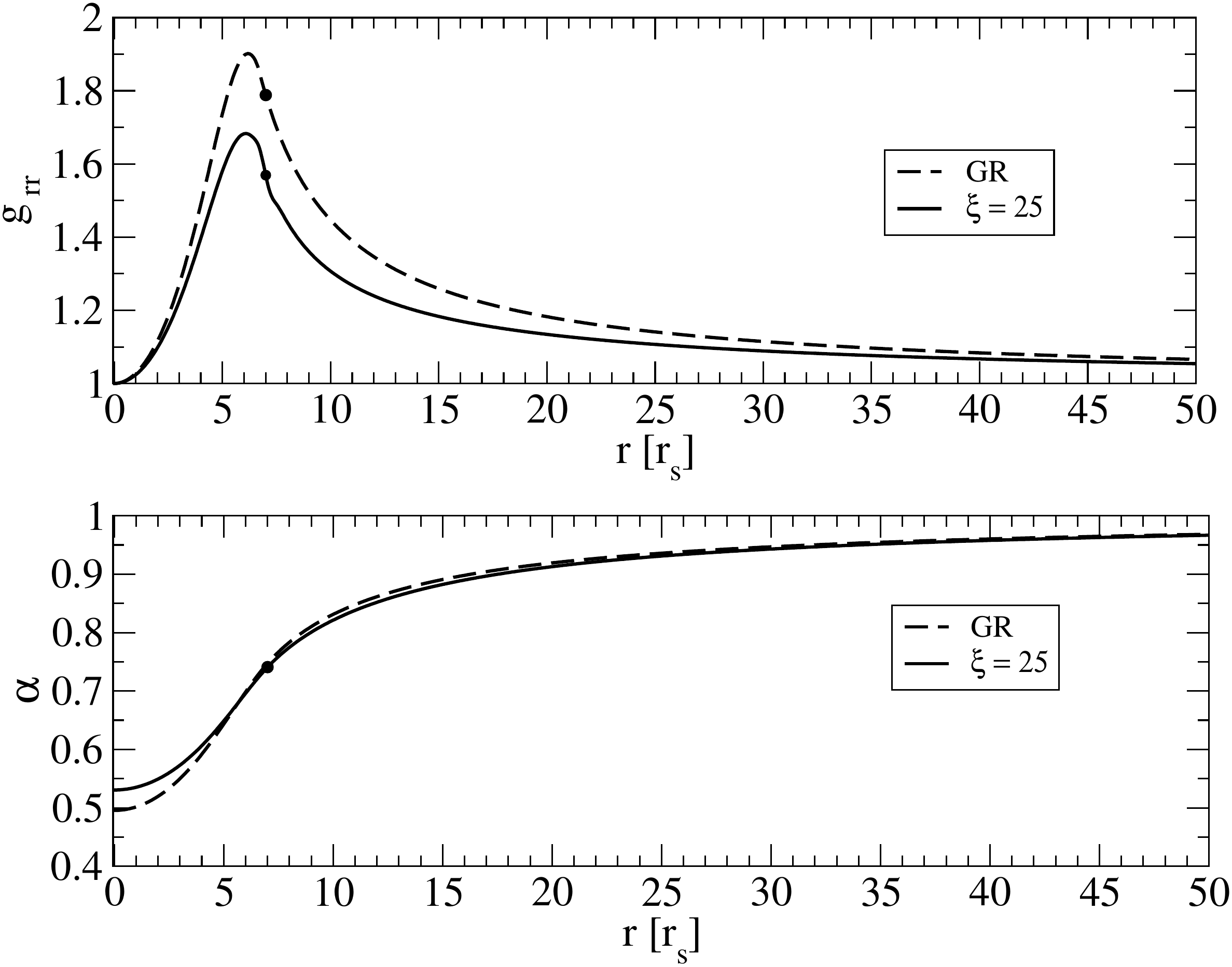}
\caption{Lapse function $\alpha(t,r)$ (lower panel) and $g_{rr}(t,r)$ (upper panel) initially (dashed line) where $\phi(0,r)\equiv 0$ and at the end of the transition to spontaneous scalarization (solid line) where $\phi(t_f,r)\neq 0$. Remarkably, the scalarized star is less compact than the initial configuration as one can appreciate from the value of the lapse at $r=0$, which is larger after the scalarization finishes. This is also reflected in the height of the peak in $g_{rr}$. The black dot indicates the values at the surface of the star (i.e. where the pressure and density vanish).
The radial coordinate is given in units of $r_s = G_{0}M_{\odot}/c^2$
.}
\label{fig:metric}
\end{figure}

\begin{figure}[!ht]
\includegraphics[width=0.5\textwidth, height=0.35\textheight]{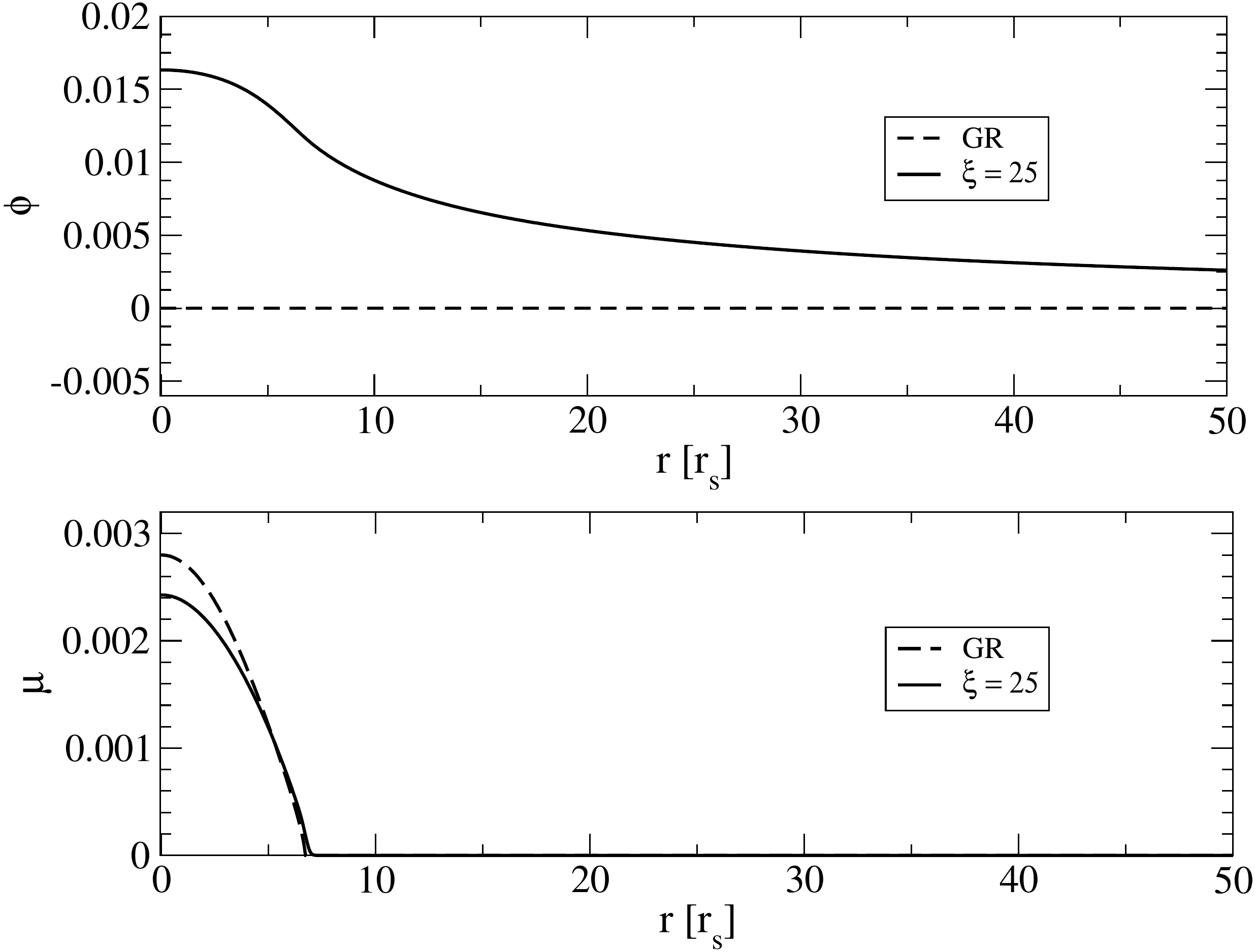}
\caption{Upper panel: Scalar field at the 
end of the scalarization process (solid line). Initially the scalar field is absent (dashed line).
Lower panel: fluid's mass density in units of $c^6/(M_\odot^2 G_0^3)\sim {3.66\times} 10^3 \mu_{\rm nuc}$. After the scalarization process the central density (solid line) becomes lower than its corresponding initial value associated with an initial unscalarized star (dashed line).}
 \label{fig:matter}
\end{figure}

\begin{figure}[!ht]
\includegraphics[width=0.5\textwidth, height=0.15\textheight]{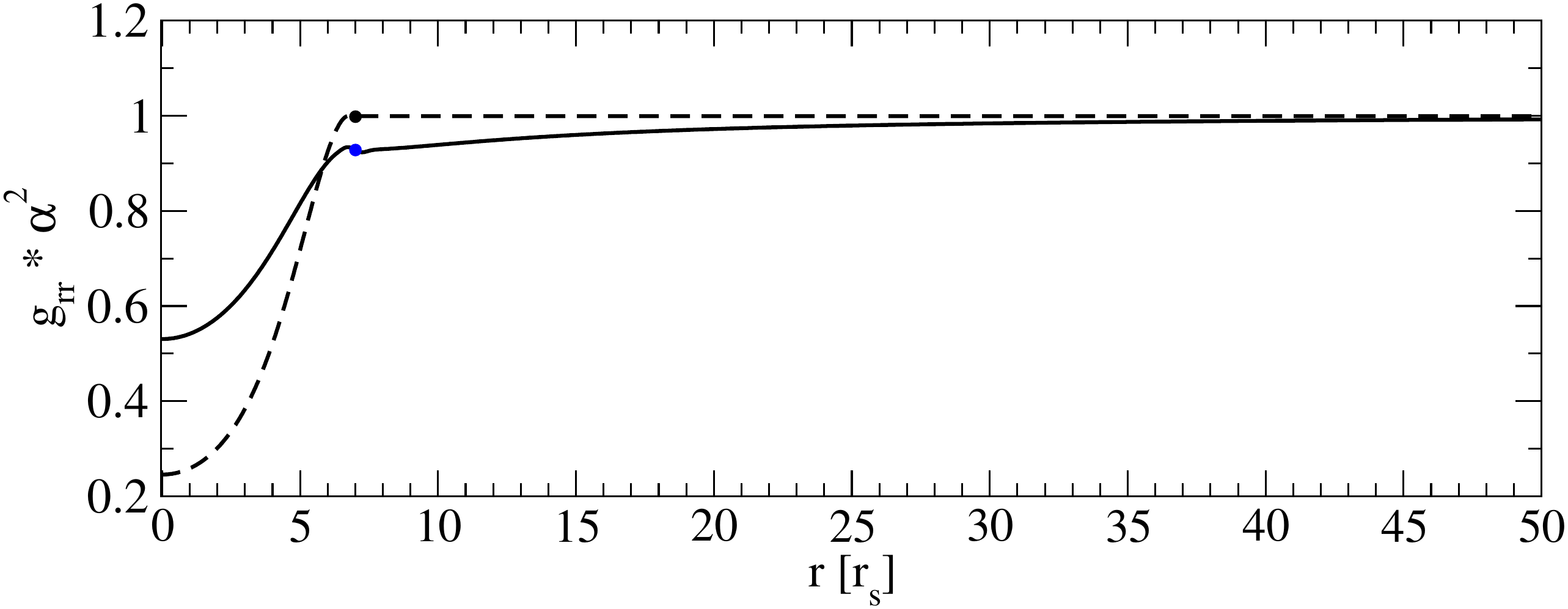}
\caption{Product $g_{rr}\times\alpha^2$, initially (dashed line) and 
at the end of the scalarization process (solid line) .
Unlike the initial (GR) configuration where this product is unity outside the star due to the absence of the scalar field (dashed line), for the scalarized solution (solid line) this product is not unity outside the star, illustrating that the spacetime is not given by the Schwarzschild solution there due to the contribution of the scalar field (the Birkhoff theorem does not apply in this case). Notwithstanding, in the asymptotic region the product does tend to unity.} 
\label{fig:metprod}
\end{figure}

Representative sequences of equilibrium models of neutron stars are plotted in Figure \ref{fig:k1527_mu_vs_mass_xis_inset} which shows different mass profiles as a function of $\mu_0$, the density at the center of the star. The location of the maximum mass indicates the critical point
separating the stable and the unstable branches towards gravitational collapse. In GR, static configurations located at the left of that point are stable, while those on the right are unstable. However, in the current analysis there also exists a point of instability (i.e. a critical density) towards SC. This critical density is 
marked with a green box
in Fig.~\ref{fig:k1527_mu_vs_mass_xis_inset}. The configurations between that density and the density leading to the maximum mass model, scalarize spontaneously and end up in one of the configurations associated with the colored lines below the black one. 
Figure~\ref{fig:metric} shows some examples of the 
metric component $g_{rr}(r)$ (upper panel) and the lapse function $\alpha(r)$ (lower panel) prior 
scalarization (dashed lines) and after scalarization (solid lines) when 
$\xi=25$. Figure \ref{fig:matter} (lower panel) depicts the corresponding densities prior (dashed line) and 
after scalarization (solid line). 
The solid lines of Figures \ref{fig:metric} and \ref{fig:matter} correspond to a configuration associated to the point (marked with a * ) in Figure 1. The upper panel of Figure \ref{fig:matter} shows the scalar 
field profile after the star becomes scalarized. Before the spontaneous scalarization initiates the scalar field is null. 
Figure \ref{fig:metprod} depicts the 
product $P:=-g_{rr}\times g_{tt}$, 
where $g_{tt}=-\alpha^2$, as a function 
of the area coordinate $r$. As we stressed, prior 
scalarization the scalar field is 
absent, thus, outside the star 
there is strict vacuum and the 
solution there is given by the 
Schwarzschild solution for which 
the product $P\equiv 1$ (solid line). 
However, at the end of scalarization 
the scalar field is present outside 
the star, and the solution is not longer the Schwarzschild solution. This is reflected by the fact that the product 
$P$ is not unity (dashed line), but only approaches that value asymptotically as the scalar field vanishes.

{\bf Dynamical evolution.}

We performed non-linear numerical evolutions in spherical symmetry of the STT-perfect-fluid system using as initial data
a static TOV configuration with the \emph{OllinSphere} code, a numerical relativity finite-difference code
for spherical symmetry.

The code implements a variant of the Baumgarte-Shapiro-Shibata-Nakamura (BSSN) evolution equations for the geometry~\cite{Alcubierre:2010is,Brown:2009dd}, which includes the contribution of the 
NMC scalar field, and for the fluid  the \emph{Valencia} formulation for relativistic hydrodynamics is implemented \cite{Font:2008fka}.
Explicit details about our numerical implementation have been reported for instance in \cite{Alcubierre:2010ea,Ruiz:2012jt}. 
We also note that
the convergence properties, gauge choices, and boundary conditions of our numerical code have
been extensively tested before in various physical systems \cite{Alcubierre:2015ipa, Alcubierre:2019qnh}. 

Under the assumption of spherical symmetry we adopt a 
conformal decomposition 
for the $3$-metric as follows,

\begin{equation}
ds^2_3 = \psi^{4}\Big[a(t, r){dr}^2 
+ r^2 b(t, r)d\Omega^2 \Big] \ ,    
\end{equation}
where $d\Omega^2 = \sin^2 \theta d\varphi^2 + d\theta^2$ is the solid angle element.
We adopt a BSSN formulation of the STT 
field equations as written 
in the form of Einstein’s equations (under spherical symmetry) 
with an effective 
EMT given by $T_{ab}$
as described below 
Eq.(\ref{eq:action_jordan}), 
which is composed by 
the three contributions 
Eqs.(\ref{eq:TabF})--(\ref{eq:Tabfluid}). Thus, the 
following quantities 
are evolved in time from the initial 
data: the metric functions $a(t, r)$ and $b(t, r)$, the conformal factor $\psi$, the trace of the extrinsic curvature $K$, the traceless part of the conformal extrinsic curvature and the radial component of the conformal connection functions. However, during the evolution $\psi$ and $b$ remain constant to their  initial values $\psi=1$ and $b=1$.

The matter fields include the contribution of both the fluid and the scalar field. 
The dynamical equations for the fluid are the relativistic Euler equation for the velocity field and for the total energy density measured by Eulerian observers, both
resulting from 
$\nabla_a T^{ab}_{\rm fluid}=0$. These equations 
can be written in {\it conservative form} following the the {\it Valencia formulation}~\cite{Font:2008fka}, which are complemented by the equation for the conservation of the baryon density (i.e. rest mass density).
For the scalar field $\phi$ we transform the second order "Klein Gordon" equation 
included below Eq.(\ref{eq:action_jordan})
into a system of two first-order evolution equations~\cite{Salgado06}.
At each time step we compute the matter sources for the effective Einstein equations. In addition to the BSSN spacetime variables and matter content, there are two more variables left undetermined, the lapse function $\alpha$, and the shift vector $\beta^r$.
For our simulations we choose for simplicity a vanishing shift throughout the evolution, while the lapse function is evolved from the initial data using the standard 1+log slicing condition: 
$\partial_t\alpha = -2\alpha K $. 
We 
perform a {\it free evolution}, and so we monitor at each time step the Hamiltonian and the momentum constraints to check the accuracy of the numerical solution, without solving the constraints at every time step, but only initially.
A detailed description 
of the specific 
equations in spherical symmetry and their 
numerical implementation will be provided in a forthcoming 
investigation where a thorough study about this subject matter will be reported, which include the collapse of a neutron star into a black hole and the radiation in the form of scalar gravitational waves associated with the NMC field $\phi$~\cite{Degollado:2021}.

\begin{figure}[!ht]
\includegraphics[width=0.5\textwidth, height=0.27\textheight]{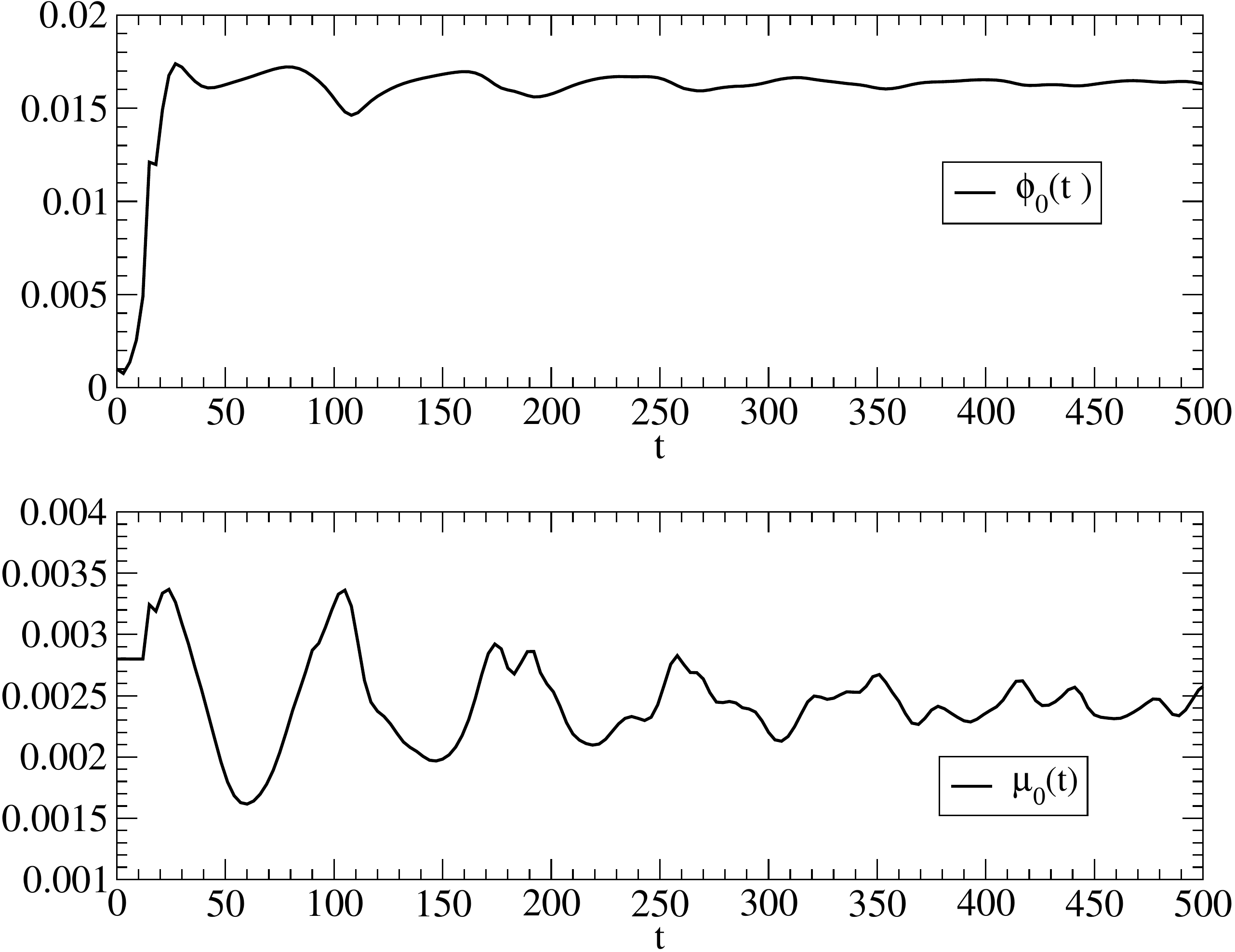}
\caption{Time evolution of the central value of the scalar field (upper panel) and central value of the total energy density of the fluid (lower panel). The system scalarizes and then oscillates around an equilibrium state.}
\label{fig:central_values}
\end{figure}

Figure~\ref{fig:central_values} depicts the evolution of the 
central value of the density 
of the star (lower panel) and the 
central value of the scalar field 
(upper panel) during the scalarzation 
process. Notice that as the 
scalarization ends, both values 
provide the central values of the 
scalarized profiles of Figure \ref{fig:matter}. Several 
physical quantities evolve 
during the scalarization process but 
for brevity we report only these two. 
In a more detailed report we 
plan to provide a more complete set of 
plots showing the evolution of 
other interesting variables.

{\bf Discussion and Outlook}
The phenomenon of spontaneous scalarization in STT allows to describe neutron stars with maximum masses that are larger than those in GR \cite{Damour93,Salgado98}. 
Beyond some critical baryon mass configurations with a non vanishing scalar field are
energetically more favorable than the corresponding configurations at the same baryon mass with 
zero scalar field.
However, scalarized neutron stars with masses larger than the maximum mass models in absence of the scalar field (i.e. those in GR) cannot be produced dynamically while keeping the baryonic mass of the star fixed since a dynamical scalarization process produces configurations with energy (i.e. gravitational mass) lower that the non scalarized counter part.
Thus, the stationary scalarized 
neutron stars with masses larger that the maximum mass models in GR described in the past (for a given equation of state) \cite{Damour93,Salgado98}, if existing in nature, must have formed following a dynamical process different from the pure spontaneous scalarization as described in this letter. Such process may include a cosmological origin or accretion of a surrounding scalar field. No doubt that the formation of scalarized neutron stars with masses 
larger than the maximum mass models 
of GR deserves further investigation as such supermassive 
scalarized neutron stars may explain 
the existence of the recently observed 
supermassive pulsars with masses larger that $2M_\odot$ without necessarily 
appealing to exotic or very stiff equations of state for the nuclear matter at high densities. While understanding this issue is a matter of 
principle, from the observational point of 
view it may well happen that such supermassive scalarized stars are already ruled out by observations in binary systems 
for requiring a large value for the NMC 
constant $\xi$. Such constraints emerge, however, when the scalar field is massless, and thus of large range, so the inclusion of a mass term for $\phi$ may help to avoid those constraints~\cite{Pretorius2016} while possibly leading to supermassive scalarized neutron stars with soft equations of state. 
This kind of EOS are presumably ruled out within the framework of GR for not reproducing the recently observed large masses in pulsars~\cite{Chamel2013,Zhou2019,Salgado94}.

{\bf Acknowledgements}
We would like to thank N. Ortiz for his comments and suggestions during the preparation of the manuscript.
This work was supported in part by
DGAPA-UNAM through grants IA101318, IN105920 and 
IN111719 and by the European Union’s Horizon 2020 research and innovation (RISE) program H2020-MSCA-RISE-2017 Grant
No. FunFiCO-777740.
\newpage

\bibliography{main.bib}

\begin{thebibliography}{53}%
\makeatletter
\providecommand \@ifxundefined [1]{%
 \@ifx{#1\undefined}
}%
\providecommand \@ifnum [1]{%
 \ifnum #1\expandafter \@firstoftwo
 \else \expandafter \@secondoftwo
 \fi
}%
\providecommand \@ifx [1]{%
 \ifx #1\expandafter \@firstoftwo
 \else \expandafter \@secondoftwo
 \fi
}%
\providecommand \natexlab [1]{#1}%
\providecommand \enquote  [1]{``#1''}%
\providecommand \bibnamefont  [1]{#1}%
\providecommand \bibfnamefont [1]{#1}%
\providecommand \citenamefont [1]{#1}%
\providecommand \href@noop [0]{\@secondoftwo}%
\providecommand \href [0]{\begingroup \@sanitize@url \@href}%
\providecommand \@href[1]{\@@startlink{#1}\@@href}%
\providecommand \@@href[1]{\endgroup#1\@@endlink}%
\providecommand \@sanitize@url [0]{\catcode `\\12\catcode `\$12\catcode
  `\&12\catcode `\#12\catcode `\^12\catcode `\_12\catcode `\%12\relax}%
\providecommand \@@startlink[1]{}%
\providecommand \@@endlink[0]{}%
\providecommand \url  [0]{\begingroup\@sanitize@url \@url }%
\providecommand \@url [1]{\endgroup\@href {#1}{\urlprefix }}%
\providecommand \urlprefix  [0]{URL }%
\providecommand \Eprint [0]{\href }%
\providecommand \doibase [0]{http://dx.doi.org/}%
\providecommand \selectlanguage [0]{\@gobble}%
\providecommand \bibinfo  [0]{\@secondoftwo}%
\providecommand \bibfield  [0]{\@secondoftwo}%
\providecommand \translation [1]{[#1]}%
\providecommand \BibitemOpen [0]{}%
\providecommand \bibitemStop [0]{}%
\providecommand \bibitemNoStop [0]{.\EOS\space}%
\providecommand \EOS [0]{\spacefactor3000\relax}%
\providecommand \BibitemShut  [1]{\csname bibitem#1\endcsname}%
\let\auto@bib@innerbib\@empty
\bibitem [{\citenamefont {Will}(tion)}]{Will2018}%
  \BibitemOpen
  \bibfield  {author} {\bibinfo {author} {\bibfnamefont {C.~M.}\ \bibnamefont
  {Will}},\ }\href@noop {} {\emph {\bibinfo {title} {Theory and experiment in
  gravitational physics}}}\ (\bibinfo  {publisher} {Cambridge University
  Press},\ \bibinfo {address} {Cambridge, U.K.},\ \bibinfo {year} {2018 (second
  edition)})\BibitemShut {NoStop}%
\bibitem [{\citenamefont {Will}(2005)}]{Will:2005va}%
  \BibitemOpen
  \bibfield  {author} {\bibinfo {author} {\bibfnamefont {C.~M.}\ \bibnamefont
  {Will}},\ }\href@noop {} {\bibfield  {journal} {\bibinfo  {journal} {Living
  Rev.Rel.}\ }\textbf {\bibinfo {volume} {9}},\ \bibinfo {pages} {3} (\bibinfo
  {year} {2005})},\ \bibinfo {note} {an update of the Living Review article
  originally published in 2001},\ \Eprint {http://arxiv.org/abs/gr-qc/0510072}
  {arXiv:gr-qc/0510072 [gr-qc]} \BibitemShut {NoStop}%
\bibitem [{\citenamefont {Berti}\ \emph {et~al.}(2015)\citenamefont {Berti}
  \emph {et~al.}}]{Berti:2015itd}%
  \BibitemOpen
  \bibfield  {author} {\bibinfo {author} {\bibfnamefont {E.}~\bibnamefont
  {Berti}} \emph {et~al.},\ }\href {\doibase 10.1088/0264-9381/32/24/243001}
  {\bibfield  {journal} {\bibinfo  {journal} {Class. Quantum Grav.}\ }\textbf
  {\bibinfo {volume} {32}},\ \bibinfo {pages} {243001} (\bibinfo {year}
  {2015})},\ \Eprint {http://arxiv.org/abs/1501.07274} {arXiv:1501.07274
  [gr-qc]} \BibitemShut {NoStop}%
\bibitem [{\citenamefont {Baker}\ \emph {et~al.}(2019)\citenamefont {Baker}
  \emph {et~al.}}]{Baker:2019gxo}%
  \BibitemOpen
  \bibfield  {author} {\bibinfo {author} {\bibfnamefont {T.}~\bibnamefont
  {Baker}} \emph {et~al.},\ }\href@noop {} {\  (\bibinfo {year} {2019})},\
  \Eprint {http://arxiv.org/abs/1908.03430} {arXiv:1908.03430 [astro-ph.CO]}
  \BibitemShut {NoStop}%
\bibitem [{\citenamefont {Burrage}\ and\ \citenamefont
  {Sakestein}(2018)}]{Burrage2018}%
  \BibitemOpen
  \bibfield  {author} {\bibinfo {author} {\bibfnamefont {C.}~\bibnamefont
  {Burrage}}\ and\ \bibinfo {author} {\bibfnamefont {J.}~\bibnamefont
  {Sakestein}},\ }\href@noop {} {\bibfield  {journal} {\bibinfo  {journal}
  {Living Rev.Rel.}\ }\textbf {\bibinfo {volume} {21}},\ \bibinfo {pages} {1}
  (\bibinfo {year} {2018})},\ \Eprint {http://arxiv.org/abs/gr-qc/1709.09071}
  {arXiv:gr-qc/1709.09071} \BibitemShut {NoStop}%
\bibitem [{\citenamefont {Joyce}\ \emph {et~al.}(2015)\citenamefont {Joyce},
  \citenamefont {Jain}, \citenamefont {Khoury},\ and\ \citenamefont
  {Trodden}}]{Joyce:2014kja}%
  \BibitemOpen
  \bibfield  {author} {\bibinfo {author} {\bibfnamefont {A.}~\bibnamefont
  {Joyce}}, \bibinfo {author} {\bibfnamefont {B.}~\bibnamefont {Jain}},
  \bibinfo {author} {\bibfnamefont {J.}~\bibnamefont {Khoury}}, \ and\ \bibinfo
  {author} {\bibfnamefont {M.}~\bibnamefont {Trodden}},\ }\href {\doibase
  10.1016/j.physrep.2014.12.002} {\bibfield  {journal} {\bibinfo  {journal}
  {Phys. Rept.}\ }\textbf {\bibinfo {volume} {568}},\ \bibinfo {pages} {1}
  (\bibinfo {year} {2015})},\ \Eprint {http://arxiv.org/abs/1407.0059}
  {arXiv:1407.0059 [astro-ph.CO]} \BibitemShut {NoStop}%
\bibitem [{\citenamefont {Martin}(2012)}]{Martin2012}%
  \BibitemOpen
  \bibfield  {author} {\bibinfo {author} {\bibfnamefont {J.}~\bibnamefont
  {Martin}},\ }\href@noop {} {\bibfield  {journal} {\bibinfo  {journal} {C. R.
  Physique}\ }\textbf {\bibinfo {volume} {13}},\ \bibinfo {pages} {566}
  (\bibinfo {year} {2012})}\BibitemShut {NoStop}%
\bibitem [{\citenamefont {Koutsoumbas}\ \emph {et~al.}(2018)\citenamefont
  {Koutsoumbas}, \citenamefont {Ntrekis}, \citenamefont {Papantonopoulos},\
  and\ \citenamefont {Saridakis}}]{Koutsoumbas2018}%
  \BibitemOpen
  \bibfield  {author} {\bibinfo {author} {\bibfnamefont {G.}~\bibnamefont
  {Koutsoumbas}}, \bibinfo {author} {\bibfnamefont {K.}~\bibnamefont
  {Ntrekis}}, \bibinfo {author} {\bibfnamefont {E.}~\bibnamefont
  {Papantonopoulos}}, \ and\ \bibinfo {author} {\bibfnamefont {E.~N.}\
  \bibnamefont {Saridakis}},\ }\href@noop {} {\bibfield  {journal} {\bibinfo
  {journal} {JCAP}\ }\textbf {\bibinfo {volume} {1802}},\ \bibinfo {pages} {3}
  (\bibinfo {year} {2018})}\BibitemShut {NoStop}%
\bibitem [{\citenamefont {Faraoni}(2004)}]{Faraoni04}%
  \BibitemOpen
  \bibfield  {author} {\bibinfo {author} {\bibfnamefont {V.}~\bibnamefont
  {Faraoni}},\ }\href@noop {} {\emph {\bibinfo {title} {Cosmology in
  Scalar-Tensor Gravity}}}\ (\bibinfo  {publisher} {Kluwer Academic
  Publishers},\ \bibinfo {address} {Dordrecht, The Netherlands},\ \bibinfo
  {year} {2004})\BibitemShut {NoStop}%
\bibitem [{\citenamefont {Tsujikawa}\ \emph {et~al.}(2008)\citenamefont
  {Tsujikawa}, \citenamefont {Uddin}, \citenamefont {Mizuno}, \citenamefont
  {Tavakol},\ and\ \citenamefont {Yokoyama}}]{Tsujikawa:2008}%
  \BibitemOpen
  \bibfield  {author} {\bibinfo {author} {\bibfnamefont {S.}~\bibnamefont
  {Tsujikawa}}, \bibinfo {author} {\bibfnamefont {K.}~\bibnamefont {Uddin}},
  \bibinfo {author} {\bibfnamefont {S.}~\bibnamefont {Mizuno}}, \bibinfo
  {author} {\bibfnamefont {R.}~\bibnamefont {Tavakol}}, \ and\ \bibinfo
  {author} {\bibfnamefont {J.}~\bibnamefont {Yokoyama}},\ }\href@noop {}
  {\bibfield  {journal} {\bibinfo  {journal} {Phys. Rev.}\ }\textbf {\bibinfo
  {volume} {D77}},\ \bibinfo {pages} {103009} (\bibinfo {year}
  {2008})}\BibitemShut {NoStop}%
\bibitem [{\citenamefont {De~Felice}\ and\ \citenamefont
  {Tsujikawa}(2010)}]{DeFelice:2010b}%
  \BibitemOpen
  \bibfield  {author} {\bibinfo {author} {\bibfnamefont {A.}~\bibnamefont
  {De~Felice}}\ and\ \bibinfo {author} {\bibfnamefont {S.}~\bibnamefont
  {Tsujikawa}},\ }\href@noop {} {\bibfield  {journal} {\bibinfo  {journal}
  {JCAP}\ }\textbf {\bibinfo {volume} {07}},\ \bibinfo {pages} {024} (\bibinfo
  {year} {2010})}\BibitemShut {NoStop}%
\bibitem [{\citenamefont {Khoury}(2013)}]{Khoury2013}%
  \BibitemOpen
  \bibfield  {author} {\bibinfo {author} {\bibfnamefont {J.}~\bibnamefont
  {Khoury}},\ }\href@noop {} {\bibfield  {journal} {\bibinfo  {journal}
  {Classical and Quant. Grav.}\ }\textbf {\bibinfo {volume} {30}},\ \bibinfo
  {pages} {214004} (\bibinfo {year} {2013})}\BibitemShut {NoStop}%
\bibitem [{\citenamefont {Burrage}\ and\ \citenamefont
  {Sakstein}(2018)}]{Burrage:2017qrf}%
  \BibitemOpen
  \bibfield  {author} {\bibinfo {author} {\bibfnamefont {C.}~\bibnamefont
  {Burrage}}\ and\ \bibinfo {author} {\bibfnamefont {J.}~\bibnamefont
  {Sakstein}},\ }\href {\doibase 10.1007/s41114-018-0011-x} {\bibfield
  {journal} {\bibinfo  {journal} {Living Rev. Rel.}\ }\textbf {\bibinfo
  {volume} {21}},\ \bibinfo {pages} {1} (\bibinfo {year} {2018})},\ \Eprint
  {http://arxiv.org/abs/1709.09071} {arXiv:1709.09071 [astro-ph.CO]}
  \BibitemShut {NoStop}%
\bibitem [{\citenamefont {Damour}\ and\ \citenamefont
  {Esposito-Far\`ese}(1992)}]{Damour92}%
  \BibitemOpen
  \bibfield  {author} {\bibinfo {author} {\bibfnamefont {T.}~\bibnamefont
  {Damour}}\ and\ \bibinfo {author} {\bibfnamefont {G.}~\bibnamefont
  {Esposito-Far\`ese}},\ }\href@noop {} {\bibfield  {journal} {\bibinfo
  {journal} {Class. Quantum Grav.}\ }\textbf {\bibinfo {volume} {9}},\ \bibinfo
  {pages} {2093} (\bibinfo {year} {1992})}\BibitemShut {NoStop}%
\bibitem [{\citenamefont {Riazuelo}\ and\ \citenamefont
  {Uzan}(2002)}]{Riazuelo2002}%
  \BibitemOpen
  \bibfield  {author} {\bibinfo {author} {\bibfnamefont {A.}~\bibnamefont
  {Riazuelo}}\ and\ \bibinfo {author} {\bibfnamefont {J.~P.}\ \bibnamefont
  {Uzan}},\ }\href@noop {} {\bibfield  {journal} {\bibinfo  {journal} {Phys.
  Rev.}\ }\textbf {\bibinfo {volume} {D66}},\ \bibinfo {pages} {023525}
  (\bibinfo {year} {2002})}\BibitemShut {NoStop}%
\bibitem [{\citenamefont {Fujii}\ and\ \citenamefont
  {Maeda}(2003)}]{Fujii2003}%
  \BibitemOpen
  \bibfield  {author} {\bibinfo {author} {\bibfnamefont {Y.}~\bibnamefont
  {Fujii}}\ and\ \bibinfo {author} {\bibfnamefont {K.}~\bibnamefont {Maeda}},\
  }\href@noop {} {\emph {\bibinfo {title} {The Scalar-Tensor Theory of
  Gravitation}}}\ (\bibinfo  {publisher} {Cambridge University Press},\
  \bibinfo {address} {Cambridge, England},\ \bibinfo {year} {2003})\BibitemShut
  {NoStop}%
\bibitem [{DES()}]{DESI}%
  \BibitemOpen
  \href@noop {} {\bibfield  {journal} {\bibinfo  {journal} {DESI collaboration:
  www.darkenergysurvey.org}\ }}\Eprint {http://arxiv.org/abs/arXiv:1611.00036}
  {arXiv:1611.00036} \BibitemShut {NoStop}%
\bibitem [{LSS()}]{LSST}%
  \BibitemOpen
  \href@noop {} {\bibinfo  {journal} {LSST telescope: www.lsst.org/lsst}\
  }\BibitemShut {NoStop}%
\bibitem [{\citenamefont {Amendola}\ \emph {et~al.}(2018)\citenamefont
  {Amendola} \emph {et~al.}}]{Amendola:2018}%
  \BibitemOpen
\bibfield  {journal} {  }\bibfield  {author} {\bibinfo {author} {\bibfnamefont
  {L.}~\bibnamefont {Amendola}} \emph {et~al.},\ }\href@noop {} {\bibfield
  {journal} {\bibinfo  {journal} {Living Rev.Rel.}\ }\textbf {\bibinfo {volume}
  {21}},\ \bibinfo {pages} {2} (\bibinfo {year} {2018})}\BibitemShut {NoStop}%
\bibitem [{EUC()}]{EUCLIDES}%
  \BibitemOpen
  \href@noop {} {\bibinfo  {journal} {EUCLIDES mission:
  http://sci.esa.int/science-e/www/area/index.cfm?fareaid=102}\ }\BibitemShut
  {NoStop}%
\bibitem [{\citenamefont {Abbott}\ \emph {et~al.}(2018)\citenamefont {Abbott}
  \emph {et~al.}}]{Abbott:2019}%
  \BibitemOpen
\bibfield  {journal} {  }\bibfield  {author} {\bibinfo {author} {\bibfnamefont
  {B.~P.}\ \bibnamefont {Abbott}} \emph {et~al.} (\bibinfo {collaboration}
  {Virgo, LIGO Scientific}),\ }\href@noop {} {\bibfield  {journal} {\bibinfo
  {journal} {Phys. Rev. Lett.}\ }\textbf {\bibinfo {volume} {120}},\ \bibinfo
  {pages} {201102} (\bibinfo {year} {2018})}\BibitemShut {NoStop}%
\bibitem [{\citenamefont {Damour}\ and\ \citenamefont
  {Esposito-Far\`ese}(1993)}]{Damour93}%
  \BibitemOpen
  \bibfield  {author} {\bibinfo {author} {\bibfnamefont {T.}~\bibnamefont
  {Damour}}\ and\ \bibinfo {author} {\bibfnamefont {G.}~\bibnamefont
  {Esposito-Far\`ese}},\ }\href@noop {} {\bibfield  {journal} {\bibinfo
  {journal} {Phys. Rev. Lett.}\ }\textbf {\bibinfo {volume} {70}},\ \bibinfo
  {pages} {2220} (\bibinfo {year} {1993})}\BibitemShut {NoStop}%
\bibitem [{\citenamefont {Damour}\ and\ \citenamefont
  {Esposito-Far\`ese}(1996)}]{Damour96}%
  \BibitemOpen
  \bibfield  {author} {\bibinfo {author} {\bibfnamefont {T.}~\bibnamefont
  {Damour}}\ and\ \bibinfo {author} {\bibfnamefont {G.}~\bibnamefont
  {Esposito-Far\`ese}},\ }\href@noop {} {\bibfield  {journal} {\bibinfo
  {journal} {Phys. Rev.}\ }\textbf {\bibinfo {volume} {D54}},\ \bibinfo {pages}
  {1474} (\bibinfo {year} {1996})}\BibitemShut {NoStop}%
\bibitem [{\citenamefont {Salgado}\ \emph {et~al.}(1998)\citenamefont
  {Salgado}, \citenamefont {Sudarsky},\ and\ \citenamefont
  {Nucamendi}}]{Salgado98}%
  \BibitemOpen
  \bibfield  {author} {\bibinfo {author} {\bibfnamefont {M.}~\bibnamefont
  {Salgado}}, \bibinfo {author} {\bibfnamefont {D.}~\bibnamefont {Sudarsky}}, \
  and\ \bibinfo {author} {\bibfnamefont {U.}~\bibnamefont {Nucamendi}},\
  }\href@noop {} {\bibfield  {journal} {\bibinfo  {journal} {Phys. Rev.}\
  }\textbf {\bibinfo {volume} {D58}},\ \bibinfo {pages} {124003} (\bibinfo
  {year} {1998})}\BibitemShut {NoStop}%
\bibitem [{\citenamefont {Novak}(1998)}]{Novak98b}%
  \BibitemOpen
  \bibfield  {author} {\bibinfo {author} {\bibfnamefont {J.}~\bibnamefont
  {Novak}},\ }\href@noop {} {\bibfield  {journal} {\bibinfo  {journal} {Phys.
  Rev.}\ }\textbf {\bibinfo {volume} {D58}},\ \bibinfo {pages} {064019}
  (\bibinfo {year} {1998})},\ \Eprint {http://arxiv.org/abs/gr-qc/9806022}
  {gr-qc/9806022} \BibitemShut {NoStop}%
\bibitem [{\citenamefont {Damour}\ and\ \citenamefont
  {Esposito-Far\`ese}(1998)}]{Damour98}%
  \BibitemOpen
  \bibfield  {author} {\bibinfo {author} {\bibfnamefont {T.}~\bibnamefont
  {Damour}}\ and\ \bibinfo {author} {\bibfnamefont {G.}~\bibnamefont
  {Esposito-Far\`ese}},\ }\href@noop {} {\bibfield  {journal} {\bibinfo
  {journal} {Phys. Rev.}\ }\textbf {\bibinfo {volume} {D58}},\ \bibinfo {pages}
  {042001} (\bibinfo {year} {1998})}\BibitemShut {NoStop}%
\bibitem [{\citenamefont {Anderson}\ and\ \citenamefont
  {Yunes}(2019)}]{Anderson2019}%
  \BibitemOpen
  \bibfield  {author} {\bibinfo {author} {\bibfnamefont {D.}~\bibnamefont
  {Anderson}}\ and\ \bibinfo {author} {\bibfnamefont {N.}~\bibnamefont
  {Yunes}},\ }\href@noop {} {\bibfield  {journal} {\bibinfo  {journal} {Class.
  Quantum Grav.}\ }\textbf {\bibinfo {volume} {36}},\ \bibinfo {pages} {165003}
  (\bibinfo {year} {2019})}\BibitemShut {NoStop}%
\bibitem [{\citenamefont {Antoniadis}\ \emph {et~al.}(2013)\citenamefont
  {Antoniadis} \emph {et~al.}}]{Antoniadis2013}%
  \BibitemOpen
  \bibfield  {author} {\bibinfo {author} {\bibfnamefont {J.}~\bibnamefont
  {Antoniadis}} \emph {et~al.},\ }\href@noop {} {\bibfield  {journal} {\bibinfo
   {journal} {Science}\ }\textbf {\bibinfo {volume} {340}},\ \bibinfo {pages}
  {1233232} (\bibinfo {year} {2013})}\BibitemShut {NoStop}%
\bibitem [{\citenamefont {Demorest}\ \emph {et~al.}(2010)\citenamefont
  {Demorest}, \citenamefont {Pennucci}, \citenamefont {Ransom}, \citenamefont
  {Roberts},\ and\ \citenamefont {Hessels}}]{Demorest2010}%
  \BibitemOpen
  \bibfield  {author} {\bibinfo {author} {\bibfnamefont {P.~B.}\ \bibnamefont
  {Demorest}}, \bibinfo {author} {\bibfnamefont {T.}~\bibnamefont {Pennucci}},
  \bibinfo {author} {\bibfnamefont {S.~M.}\ \bibnamefont {Ransom}}, \bibinfo
  {author} {\bibfnamefont {M.~S.~E.}\ \bibnamefont {Roberts}}, \ and\ \bibinfo
  {author} {\bibfnamefont {J.~W.~T.}\ \bibnamefont {Hessels}},\ }\href@noop {}
  {\bibfield  {journal} {\bibinfo  {journal} {Nature}\ }\textbf {\bibinfo
  {volume} {467}},\ \bibinfo {pages} {1081} (\bibinfo {year}
  {2010})}\BibitemShut {NoStop}%
\bibitem [{\citenamefont {Cromartie}\ \emph {et~al.}(80 2)\citenamefont
  {Cromartie} \emph {et~al.}}]{Cromartie2019}%
  \BibitemOpen
  \bibfield  {author} {\bibinfo {author} {\bibfnamefont {H.~T.}\ \bibnamefont
  {Cromartie}} \emph {et~al.},\ }\href@noop {} {\bibfield  {journal} {\bibinfo
  {journal} {Nature Astronomy}\ } (\bibinfo {year}
  {doi:10.01038/s41550-019-0880-2})}\BibitemShut {NoStop}%
\bibitem [{\citenamefont {Linares}\ \emph {et~al.}(2018)\citenamefont
  {Linares}, \citenamefont {Shahbaz},\ and\ \citenamefont
  {Casares}}]{Linares2018}%
  \BibitemOpen
  \bibfield  {author} {\bibinfo {author} {\bibfnamefont {M.}~\bibnamefont
  {Linares}}, \bibinfo {author} {\bibfnamefont {T.}~\bibnamefont {Shahbaz}}, \
  and\ \bibinfo {author} {\bibfnamefont {J.}~\bibnamefont {Casares}},\
  }\href@noop {} {\bibfield  {journal} {\bibinfo  {journal} {Astrophys. J}\
  }\textbf {\bibinfo {volume} {859}},\ \bibinfo {pages} {1} (\bibinfo {year}
  {2018})}\BibitemShut {NoStop}%
\bibitem [{\citenamefont {Chamel}\ \emph {et~al.}(2013)\citenamefont {Chamel},
  \citenamefont {Haensel}, \citenamefont {Zdunik},\ and\ \citenamefont
  {Fantina}}]{Chamel2013}%
  \BibitemOpen
  \bibfield  {author} {\bibinfo {author} {\bibfnamefont {N.}~\bibnamefont
  {Chamel}}, \bibinfo {author} {\bibfnamefont {P.}~\bibnamefont {Haensel}},
  \bibinfo {author} {\bibfnamefont {J.~L.}\ \bibnamefont {Zdunik}}, \ and\
  \bibinfo {author} {\bibfnamefont {A.~F.}\ \bibnamefont {Fantina}},\
  }\href@noop {} {\bibfield  {journal} {\bibinfo  {journal} {IJMPE}\ }\textbf
  {\bibinfo {volume} {22}},\ \bibinfo {pages} {1330018} (\bibinfo {year}
  {2013})}\BibitemShut {NoStop}%
\bibitem [{\citenamefont {Zhou}\ and\ \citenamefont {Chen}(2019)}]{Zhou2019}%
  \BibitemOpen
  \bibfield  {author} {\bibinfo {author} {\bibfnamefont {Y.}~\bibnamefont
  {Zhou}}\ and\ \bibinfo {author} {\bibfnamefont {L.~W.}\ \bibnamefont
  {Chen}},\ }\href@noop {} {\bibfield  {journal} {\bibinfo  {journal}
  {Astrophys. J}\ }\textbf {\bibinfo {volume} {886}},\ \bibinfo {pages} {52}
  (\bibinfo {year} {2019})}\BibitemShut {NoStop}%
\bibitem [{\citenamefont {Salgado}\ \emph {et~al.}(1994)\citenamefont
  {Salgado}, \citenamefont {Bonazzola}, \citenamefont {Gourgoulhon},\ and\
  \citenamefont {Haensel}}]{Salgado94}%
  \BibitemOpen
  \bibfield  {author} {\bibinfo {author} {\bibfnamefont {M.}~\bibnamefont
  {Salgado}}, \bibinfo {author} {\bibfnamefont {S.}~\bibnamefont {Bonazzola}},
  \bibinfo {author} {\bibfnamefont {E.}~\bibnamefont {Gourgoulhon}}, \ and\
  \bibinfo {author} {\bibfnamefont {P.}~\bibnamefont {Haensel}},\ }\href@noop
  {} {\bibfield  {journal} {\bibinfo  {journal} {Astron. Astrophys.}\ }\textbf
  {\bibinfo {volume} {291}},\ \bibinfo {pages} {155} (\bibinfo {year}
  {1994})}\BibitemShut {NoStop}%
\bibitem [{\citenamefont {Ortiz}\ and\ \citenamefont
  {Mendes}(2016)}]{Ortiz2016}%
  \BibitemOpen
  \bibfield  {author} {\bibinfo {author} {\bibfnamefont {N.}~\bibnamefont
  {Ortiz}}\ and\ \bibinfo {author} {\bibfnamefont {R.}~\bibnamefont {Mendes}},\
  }\href@noop {} {\bibfield  {journal} {\bibinfo  {journal} {Phys. Rev.}\
  }\textbf {\bibinfo {volume} {D93}},\ \bibinfo {pages} {124035} (\bibinfo
  {year} {2016})},\ \Eprint {http://arxiv.org/abs/gr-qc/1604.04175}
  {gr-qc/1604.04175} \BibitemShut {NoStop}%
\bibitem [{\citenamefont {Damour}\ and\ \citenamefont
  {Esposito-Farese}(1993)}]{Damour:1993hw}%
  \BibitemOpen
  \bibfield  {author} {\bibinfo {author} {\bibfnamefont {T.}~\bibnamefont
  {Damour}}\ and\ \bibinfo {author} {\bibfnamefont {G.}~\bibnamefont
  {Esposito-Farese}},\ }\href {\doibase 10.1103/PhysRevLett.70.2220} {\bibfield
   {journal} {\bibinfo  {journal} {Phys. Rev. Lett.}\ }\textbf {\bibinfo
  {volume} {70}},\ \bibinfo {pages} {2220} (\bibinfo {year}
  {1993})}\BibitemShut {NoStop}%
\bibitem [{\citenamefont {Abbott}\ \emph
  {et~al.}(2017{\natexlab{a}})\citenamefont {Abbott} \emph
  {et~al.}}]{Abbott:2017a}%
  \BibitemOpen
  \bibfield  {author} {\bibinfo {author} {\bibfnamefont {B.~P.}\ \bibnamefont
  {Abbott}} \emph {et~al.} (\bibinfo {collaboration} {Virgo, LIGO
  Scientific}),\ }\href@noop {} {\bibfield  {journal} {\bibinfo  {journal}
  {Phys. Rev. Lett.}\ }\textbf {\bibinfo {volume} {119}},\ \bibinfo {pages}
  {161101} (\bibinfo {year} {2017}{\natexlab{a}})}\BibitemShut {NoStop}%
\bibitem [{\citenamefont {Abbott}\ \emph
  {et~al.}(2017{\natexlab{b}})\citenamefont {Abbott} \emph
  {et~al.}}]{Abbott:2017b}%
  \BibitemOpen
  \bibfield  {author} {\bibinfo {author} {\bibfnamefont {B.~P.}\ \bibnamefont
  {Abbott}} \emph {et~al.} (\bibinfo {collaboration} {Virgo, Fermi-GBM,
  INTEGRAL, LIGO Scientific collaborations}),\ }\href@noop {} {\bibfield
  {journal} {\bibinfo  {journal} {Astrophys J.}\ }\textbf {\bibinfo {volume}
  {848}},\ \bibinfo {pages} {L13} (\bibinfo {year}
  {2017}{\natexlab{b}})}\BibitemShut {NoStop}%
\bibitem [{\citenamefont {Ezquiaga}\ and\ \citenamefont
  {Zumalac\'arregui}(2017)}]{Ezquiaga:2017}%
  \BibitemOpen
  \bibfield  {author} {\bibinfo {author} {\bibfnamefont {J.~M.}\ \bibnamefont
  {Ezquiaga}}\ and\ \bibinfo {author} {\bibfnamefont {M.}~\bibnamefont
  {Zumalac\'arregui}},\ }\href@noop {} {\bibfield  {journal} {\bibinfo
  {journal} {Phys. Rev. Lett.}\ }\textbf {\bibinfo {volume} {119}},\ \bibinfo
  {pages} {251304} (\bibinfo {year} {2017})}\BibitemShut {NoStop}%
\bibitem [{\citenamefont {Baker}\ \emph {et~al.}(2017)\citenamefont {Baker},
  \citenamefont {Bellini}, \citenamefont {Ferreira}, \citenamefont {Lagos},
  \citenamefont {Noller},\ and\ \citenamefont {Sawicki}}]{Baker:2017}%
  \BibitemOpen
  \bibfield  {author} {\bibinfo {author} {\bibfnamefont {T.}~\bibnamefont
  {Baker}}, \bibinfo {author} {\bibfnamefont {E.}~\bibnamefont {Bellini}},
  \bibinfo {author} {\bibfnamefont {P.~G.}\ \bibnamefont {Ferreira}}, \bibinfo
  {author} {\bibfnamefont {M.}~\bibnamefont {Lagos}}, \bibinfo {author}
  {\bibfnamefont {J.}~\bibnamefont {Noller}}, \ and\ \bibinfo {author}
  {\bibfnamefont {I.}~\bibnamefont {Sawicki}},\ }\href@noop {} {\bibfield
  {journal} {\bibinfo  {journal} {Phys. Rev. Lett.}\ }\textbf {\bibinfo
  {volume} {119}},\ \bibinfo {pages} {251301} (\bibinfo {year}
  {2017})}\BibitemShut {NoStop}%
\bibitem [{\citenamefont {Sakstein}\ and\ \citenamefont
  {Jain}(2017)}]{Sakstein:2017}%
  \BibitemOpen
  \bibfield  {author} {\bibinfo {author} {\bibfnamefont {J.}~\bibnamefont
  {Sakstein}}\ and\ \bibinfo {author} {\bibfnamefont {B.}~\bibnamefont
  {Jain}},\ }\href@noop {} {\bibfield  {journal} {\bibinfo  {journal} {Phys.
  Rev. Lett.}\ }\textbf {\bibinfo {volume} {119}},\ \bibinfo {pages} {251303}
  (\bibinfo {year} {2017})}\BibitemShut {NoStop}%
\bibitem [{\citenamefont {Ramazanoglu}\ and\ \citenamefont
  {Pretoius}(2016)}]{Pretorius2016}%
  \BibitemOpen
  \bibfield  {author} {\bibinfo {author} {\bibfnamefont {F.~M.}\ \bibnamefont
  {Ramazanoglu}}\ and\ \bibinfo {author} {\bibfnamefont {F.}~\bibnamefont
  {Pretoius}},\ }\href@noop {} {\bibfield  {journal} {\bibinfo  {journal}
  {Phys. Rev. D}\ }\textbf {\bibinfo {volume} {93}},\ \bibinfo {pages} {064005}
  (\bibinfo {year} {2016})}\BibitemShut {NoStop}%
\bibitem [{\citenamefont {Ruiz}\ \emph {et~al.}(2012)\citenamefont {Ruiz},
  \citenamefont {Degollado}, \citenamefont {Alcubierre}, \citenamefont
  {N\'u{\~n}ez},\ and\ \citenamefont {Salgado}}]{Ruiz:2012jt}%
  \BibitemOpen
  \bibfield  {author} {\bibinfo {author} {\bibfnamefont {M.}~\bibnamefont
  {Ruiz}}, \bibinfo {author} {\bibfnamefont {J.~C.}\ \bibnamefont {Degollado}},
  \bibinfo {author} {\bibfnamefont {M.}~\bibnamefont {Alcubierre}}, \bibinfo
  {author} {\bibfnamefont {D.}~\bibnamefont {N\'u{\~n}ez}}, \ and\ \bibinfo
  {author} {\bibfnamefont {M.}~\bibnamefont {Salgado}},\ }\href {\doibase
  10.1103/PhysRevD.86.104044} {\bibfield  {journal} {\bibinfo  {journal}
  {Phys.Rev.}\ }\textbf {\bibinfo {volume} {D86}},\ \bibinfo {pages} {104044}
  (\bibinfo {year} {2012})},\ \Eprint {http://arxiv.org/abs/1207.6142}
  {arXiv:1207.6142 [gr-qc]} \BibitemShut {NoStop}%
\bibitem [{Note1()}]{Note1}%
  \BibitemOpen
  \bibinfo {note} {In Sec.~VI of\cite {Damour96} a STT similar to the one
  presented here is considered. In order to match notations their scalar field
  $\Phi $ and ours $\phi $ are related by $\Phi =\protect \sqrt {\kappa }\phi
  $. In which case, the NMC constant $\xi $ coincide.}\BibitemShut {Stop}%
\bibitem [{\citenamefont {Freire}\ \emph {et~al.}(2012)\citenamefont {Freire}
  \emph {et~al.}}]{Freire2012}%
  \BibitemOpen
  \bibfield  {author} {\bibinfo {author} {\bibfnamefont {P.}~\bibnamefont
  {Freire}} \emph {et~al.},\ }\href@noop {} {\bibfield  {journal} {\bibinfo
  {journal} {Mon. Not. R. Asron. Soc.}\ }\textbf {\bibinfo {volume} {423}},\
  \bibinfo {pages} {3328} (\bibinfo {year} {2012})}\BibitemShut {NoStop}%
\bibitem [{\citenamefont {Alcubierre}\ and\ \citenamefont
  {Mendez}(2011)}]{Alcubierre:2010is}%
  \BibitemOpen
  \bibfield  {author} {\bibinfo {author} {\bibfnamefont {M.}~\bibnamefont
  {Alcubierre}}\ and\ \bibinfo {author} {\bibfnamefont {M.~D.}\ \bibnamefont
  {Mendez}},\ }\href {\doibase 10.1007/s10714-011-1202-x} {\bibfield  {journal}
  {\bibinfo  {journal} {Gen.Rel.Grav.}\ }\textbf {\bibinfo {volume} {43}},\
  \bibinfo {pages} {2769} (\bibinfo {year} {2011})},\ \Eprint
  {http://arxiv.org/abs/1010.4013} {arXiv:1010.4013 [gr-qc]} \BibitemShut
  {NoStop}%
\bibitem [{\citenamefont {Brown}(2009)}]{Brown:2009dd}%
  \BibitemOpen
  \bibfield  {author} {\bibinfo {author} {\bibfnamefont {J.~D.}\ \bibnamefont
  {Brown}},\ }\href {\doibase 10.1103/PhysRevD.79.104029} {\bibfield  {journal}
  {\bibinfo  {journal} {Phys. Rev.}\ }\textbf {\bibinfo {volume} {D79}},\
  \bibinfo {pages} {104029} (\bibinfo {year} {2009})},\ \Eprint
  {http://arxiv.org/abs/0902.3652} {arXiv:0902.3652 [gr-qc]} \BibitemShut
  {NoStop}%
\bibitem [{\citenamefont {Font}(2008)}]{Font:2008fka}%
  \BibitemOpen
  \bibfield  {author} {\bibinfo {author} {\bibfnamefont {J.~A.}\ \bibnamefont
  {Font}},\ }\href {\doibase 10.12942/lrr-2008-7} {\bibfield  {journal}
  {\bibinfo  {journal} {Living Rev.Rel.}\ }\textbf {\bibinfo {volume} {11}},\
  \bibinfo {pages} {7} (\bibinfo {year} {2008})}\BibitemShut {NoStop}%
\bibitem [{\citenamefont {Alcubierre}\ \emph {et~al.}(2010)\citenamefont
  {Alcubierre}, \citenamefont {Degollado}, \citenamefont {N\'u{\~n}ez},
  \citenamefont {Ruiz},\ and\ \citenamefont {Salgado}}]{Alcubierre:2010ea}%
  \BibitemOpen
  \bibfield  {author} {\bibinfo {author} {\bibfnamefont {M.}~\bibnamefont
  {Alcubierre}}, \bibinfo {author} {\bibfnamefont {J.~C.}\ \bibnamefont
  {Degollado}}, \bibinfo {author} {\bibfnamefont {D.}~\bibnamefont
  {N\'u{\~n}ez}}, \bibinfo {author} {\bibfnamefont {M.}~\bibnamefont {Ruiz}}, \
  and\ \bibinfo {author} {\bibfnamefont {M.}~\bibnamefont {Salgado}},\ }\href
  {\doibase 10.1103/PhysRevD.81.124018} {\bibfield  {journal} {\bibinfo
  {journal} {Phys. Rev.}\ }\textbf {\bibinfo {volume} {D81}},\ \bibinfo {pages}
  {124018} (\bibinfo {year} {2010})},\ \Eprint {http://arxiv.org/abs/1003.4767}
  {arXiv:1003.4767 [gr-qc]} \BibitemShut {NoStop}%
\bibitem [{\citenamefont {Alcubierre}\ \emph {et~al.}(2015)\citenamefont
  {Alcubierre}, \citenamefont {de~la Macorra}, \citenamefont {Diez-Tejedor},\
  and\ \citenamefont {Torres}}]{Alcubierre:2015ipa}%
  \BibitemOpen
  \bibfield  {author} {\bibinfo {author} {\bibfnamefont {M.}~\bibnamefont
  {Alcubierre}}, \bibinfo {author} {\bibfnamefont {A.}~\bibnamefont {de~la
  Macorra}}, \bibinfo {author} {\bibfnamefont {A.}~\bibnamefont
  {Diez-Tejedor}}, \ and\ \bibinfo {author} {\bibfnamefont {J.~M.}\
  \bibnamefont {Torres}},\ }\href {\doibase 10.1103/PhysRevD.92.063508}
  {\bibfield  {journal} {\bibinfo  {journal} {Phys. Rev.}\ }\textbf {\bibinfo
  {volume} {D92}},\ \bibinfo {pages} {063508} (\bibinfo {year} {2015})},\
  \Eprint {http://arxiv.org/abs/1501.06918} {arXiv:1501.06918 [gr-qc]}
  \BibitemShut {NoStop}%
\bibitem [{\citenamefont {Alcubierre}\ \emph {et~al.}(2019)\citenamefont
  {Alcubierre}, \citenamefont {Barranco}, \citenamefont {Bernal}, \citenamefont
  {Degollado}, \citenamefont {Diez-Tejedor}, \citenamefont {Megevand},
  \citenamefont {Núñez},\ and\ \citenamefont {Sarbach}}]{Alcubierre:2019qnh}%
  \BibitemOpen
  \bibfield  {author} {\bibinfo {author} {\bibfnamefont {M.}~\bibnamefont
  {Alcubierre}}, \bibinfo {author} {\bibfnamefont {J.}~\bibnamefont
  {Barranco}}, \bibinfo {author} {\bibfnamefont {A.}~\bibnamefont {Bernal}},
  \bibinfo {author} {\bibfnamefont {J.~C.}\ \bibnamefont {Degollado}}, \bibinfo
  {author} {\bibfnamefont {A.}~\bibnamefont {Diez-Tejedor}}, \bibinfo {author}
  {\bibfnamefont {M.}~\bibnamefont {Megevand}}, \bibinfo {author}
  {\bibfnamefont {D.}~\bibnamefont {Núñez}}, \ and\ \bibinfo {author}
  {\bibfnamefont {O.}~\bibnamefont {Sarbach}},\ }\href {\doibase
  10.1088/1361-6382/ab4726} {\bibfield  {journal} {\bibinfo  {journal} {Class.
  Quant. Grav.}\ }\textbf {\bibinfo {volume} {36}},\ \bibinfo {pages} {215013}
  (\bibinfo {year} {2019})},\ \Eprint {http://arxiv.org/abs/1906.08959}
  {arXiv:1906.08959 [gr-qc]} \BibitemShut {NoStop}%
\bibitem [{\citenamefont {Salgado}(2006)}]{Salgado06}%
  \BibitemOpen
  \bibfield  {author} {\bibinfo {author} {\bibfnamefont {M.}~\bibnamefont
  {Salgado}},\ }\href@noop {} {\bibfield  {journal} {\bibinfo  {journal}
  {Class. Quantum Grav.}\ }\textbf {\bibinfo {volume} {23}},\ \bibinfo {pages}
  {4719} (\bibinfo {year} {2006})},\ \Eprint
  {http://arxiv.org/abs/gr-qc/0509001} {gr-qc/0509001} \BibitemShut {NoStop}%
\bibitem [{\citenamefont {Degollado}\ \emph {et~al.}()\citenamefont
  {Degollado}, \citenamefont {Salgado},\ and\ \citenamefont
  {Alcubierre}}]{Degollado:2021}%
  \BibitemOpen
  \bibfield  {author} {\bibinfo {author} {\bibfnamefont {J.~C.}\ \bibnamefont
  {Degollado}}, \bibinfo {author} {\bibfnamefont {M.}~\bibnamefont {Salgado}},
  \ and\ \bibinfo {author} {\bibfnamefont {M.}~\bibnamefont {Alcubierre}},\
  }\href@noop {} {\bibinfo  {journal} {(in preparation)}\ }\BibitemShut
  {NoStop}%
\end{thebibliography}%

\end{document}